# Variability in the Atmosphere of the Hot Giant Planet HAT-P-7b

D. J. Armstrong[1,2*], E. de Mooij[2], J. Barstow[3], H. P. Osborn[1], J. Blake[1], N. Fereshteh Saniee[1]

1 Department of Physics, University of Warwick, Gibbet Hill Road, Coventry CV4 7AL
2 ARC, School of Mathematics & Physics, Queen's University Belfast, University Road, Belfast BT7 1NN, UK
3 Astrophysics Group, Department of Physics and Astronomy, University College London, London, NW1 2PS, UK

## Summary

As an exoplanet orbits its host star it reflects and emits light, forming a distinctive phase curve[1,2]. By observing this light, we can study the atmosphere and surface of distant planets. The planets in our Solar System show a wide range of atmospheric phenomena, with stable wind patterns, changing storms, and evolving anomalies. Brown dwarfs also exhibit atmospheric variability[3,4]. Such temporal variability in the atmosphere of a giant exoplanet has not to date been observed. HAT-P-7 b is an exoplanet with a known offset in the peak of its phase curve[5,6]. Here we present variations in the peak offset ranging between $-0.086^{+0.033}_{-0.033}$ to $0.143^{+0.040}_{-0.037}$ in phase, implying that the peak brightness repeatedly shifts from one side of the planet's substellar point to the other. The variability occurs on a timescale of tens to hundreds of days. These shifts in brightness are indicative of variability in the planet's atmosphere, and result from a changing balance of thermal emission and reflected flux from the planet's dayside. We suggest that variation in wind speed in the planetary atmosphere, leading to variable cloud coverage on the dayside and a changing energy balance, is capable of explaining the observed variation.

## Main Text

HAT-P-7 b is a Hot Jupiter with radius 1.4 $R_{Jupiter}$ which transits its host star with period 2.20 days[7]. It is extremely hot, with a dayside brightness temperature of 2860K, and equilibrium temperature of 2200K[6]. It was continuously observed for 4 years by the *Kepler* satellite[8] at optical wavelengths. HAT-P-7 b has also been intensively observed at Infrared (IR) wavelengths with the *Spitzer* satellite[9,10].

Both the optical and IR phase curves of HAT-P-7 b have been studied previously[5,6,9,11-13]. The optical phase curve exhibits a significant fraction of thermal emission (potentially as high as 77%[6]), due to the high temperature of the planet. The optical phase curve also presents a known eastward shift on average[6,14], which in a thermal emission dominated phase curve could be caused by the underlying global circulation pattern created by planetary scale Rossby waves arising from day-nightside temperature differences[15]. Large scale weather in the atmosphere of Hot Jupiters is theoretically expected[16-18], however no previous searches for variations in the optical phase curve parameters with time have been made. IR variability is predicted at the level of a percent[19] and has likewise not been observed. Previous work[9] found a marginally significant increase in secondary eclipse depth at 3.6μm of 59% between measurements, but this was attributed to differences in the analysis technique. Spitzer observations of multiple eclipses spaced at significant intervals have been used to put an upper limit on the variability of HD189733b at <2.7%[20,21], and have detected eclipse depth changes in the super-Earth 55 Cancri e[22]. However, the nature of those observations prevented continuous tracking of the variability.

We use 4 years of public *Kepler* data of the HAT-P-7 system. We detrend the lightcurve for instrumental effects (see Methods), and then combine the observations using a sliding bin of 10 orbits using published ephemeris[6]. Subsequently, we fit each of the combined phase-curves using a model describing both planetary and stellar effects (see Methods for details). The model combines ellipsoidal variation and Doppler beaming on the star, the planetary occultation, planetary brightness modelled as a Lambertian sphere with an offset, and a previously detected cosine third harmonic[6]. Binning over 10 orbits allows us to build up significance for each fit through gaining more datapoints, but averages out variations on timescales shorter than 10x the planetary period. This also removes stellar or instrumental variability on timescales significantly shorter than the bin size. Each successive bin starts one orbit later than the previous bin. As such, successive fit parameters are not fully independent, and hence care must be taken when assessing significance. Independent fit parameters produced from discrete bins are shown in Supplementary Figure 1. We fix the planet mass and associated parameters to published values[6], and fit for the planetary brightness amplitude $A_p$, occultation depth $F_{ecl}$ and peak brightness offset $\Theta$ for each bin of 10 planetary orbits. To assess the reliability of the measurements and determine the uncertainties, we use 4 separate methods, including a 'prayer bead' residual permutation method[23] and MCMC, as described in the Methods section.

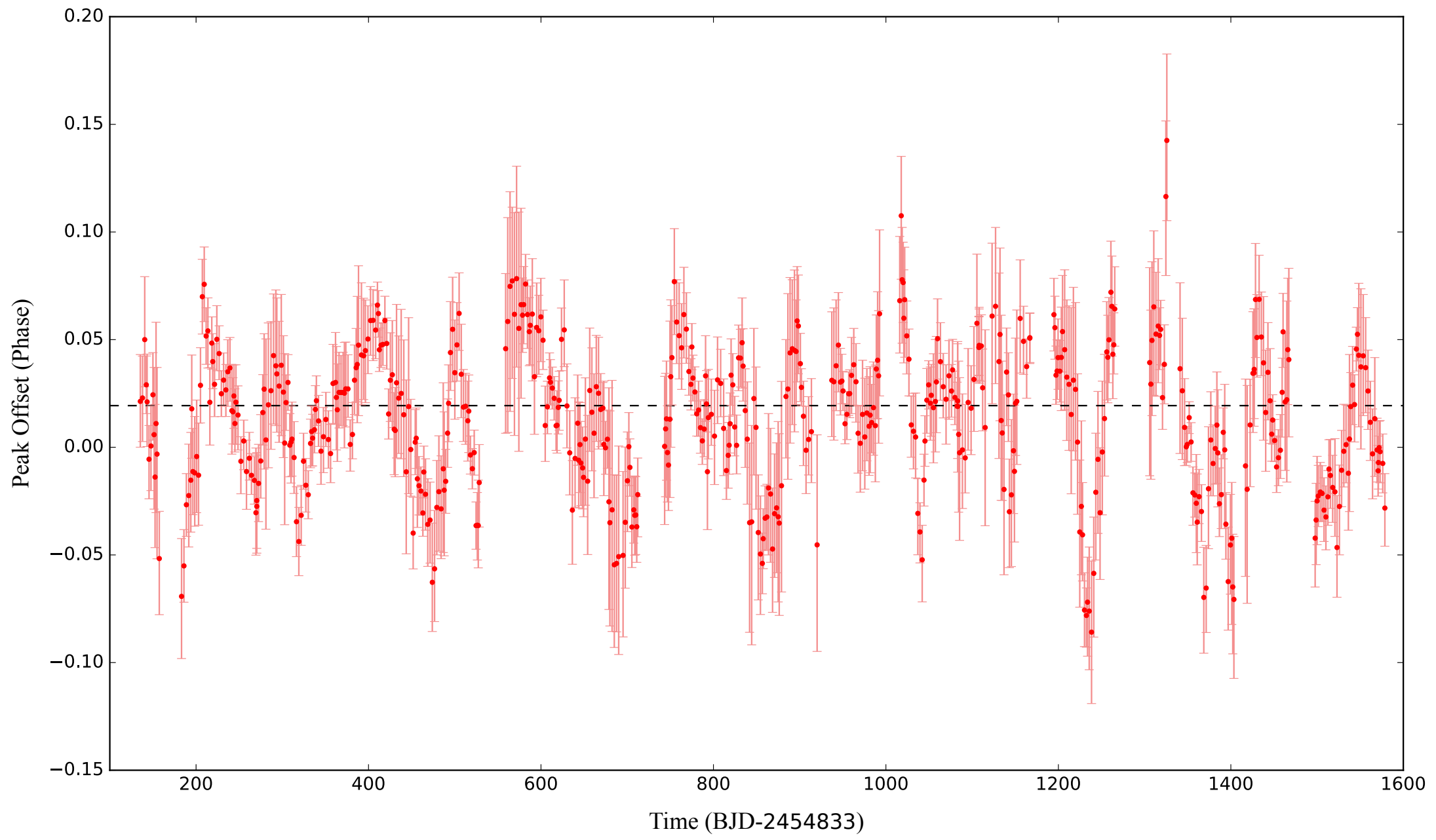

**Figure 1: Variation in peak offset of HAT-P-7 b phase curve with time.** Derived from fitting to a sliding bin of 10 planetary orbits. The horizontal black line shows the value obtained by fitting the entire dataset simultaneously. Negative offsets correspond to a movement towards the morning side of the planet, and are associated with an increased proportion of reflected light in the *Kepler* bandpass. Positive offsets represent movement to the afternoon side, and increased thermal emission in the bandpass. Error bars are 1σ errors on the fit parameters. BJD = Barycentric Julian Date.

Clear variation is seen in $\Theta$ (Figure 1). The variation spans $-0.086^{+0.033}_{-0.033}$ to $0.143^{+0.040}_{-0.037}$ in phase, with standard deviation 0.033, and an average error on the fits of 0.022 in phase. There is also marginal variation in $A_p$, with a standard deviation of 14.6ppm, but this is likely the result of systematic noise biasing the fits (see injection tests in Methods). The derived value of $A_p$ is susceptible to systematic noise occurring near the transit; $\Theta$ is more robust, as to change the peak location noise must itself have an amplitude larger than the peak. The occultation depth $F_{ecl}$ does not vary within our errors (Supplementary Figure 1).

Individual fit values are given in the Supplementary Information. Example phase curves and models are shown in Figure 2, and further examples can be found in Supplementary Figure 2. Since the *Kepler* data needs to be detrended to correct for instrument effects, we use different detrending techniques to assess the reliability of our observations. We also test for systematic biases by injecting a non-variable HAT-P-7 b phase curve into the lightcurves of two other stars. Each of these tests confirm the variations in $\Theta$, and are described in the Methods.

Weather in HAT-P-7 b's atmosphere would be expected to contribute to both the observed shifts in $\Theta$ and $A_p$ & $F_{ecl}$. While we do not observe variability in $A_p$ or $F_{ecl}$, we can limit the variations in $A_p$ to ±59% (3 standard deviations of our fit parameters), and those in $F_{ecl}$ to ±51%. This leaves a large potential for variability. Significant temperature differences could be expected if, for example, the circulation efficiency transporting heat to the planet's nightside were particularly variable, causing energy to build up on the planet's dayside.

It is important to consider what physical scenarios can lead to such significant changes in peak offset. Here we implement a published semi-analytical model[24] to gain an understanding of the origins of the observed variation (see reference for full description). The model depends primarily on the planet's Bond Albedo $A_B$, heat redistribution efficiency $\varepsilon$, cloud reflectivity boosting factor $\kappa$ and condensation temperature $T_C$. These parameters and their effects are discussed in the Supplementary Information. Both thermal emission and reflected light are included. We assume positive values for ε corresponding to superrotating winds on the planet. This assumption is supported by current Hot Jupiter atmosphere models[15]. The model allows the calculation of brightness maps on the planetary surface, and typical maps giving a positive and negative $\Theta$ are shown in Figure 3. We discuss the conditions that could lead to these cases below.

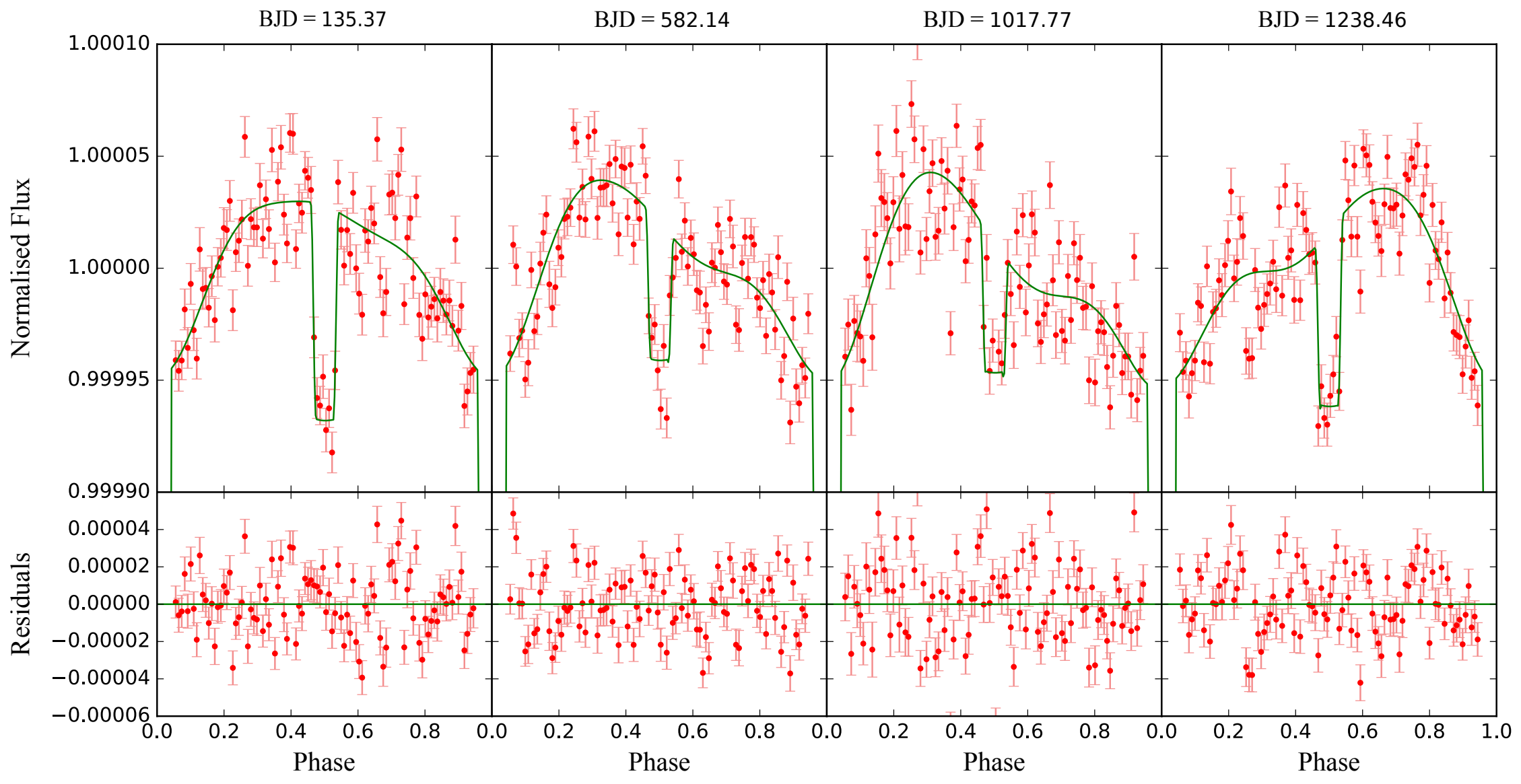

**Figure 2: Individual phase curves, with best fitting model**. The best fitting peak offsets are, from left to right, 0.021, 0.076, 0.108, -0.086. For clarity, the data is binned to 0.01 in phase before plotting, however the model was fit to the unbinned data. The lower panels show the binned residuals after the best fitting model is removed. Error bars are the standard error on the mean of the datapoints contributing to each bin.

We calculate a grid of models in $A_B$ [0-0.6]- $\varepsilon$ [0-20]- $\kappa$ [1-40] space. For each model, we fit the offset Lambertian phase curve which we applied to the data above, extracting the corresponding $A_p$, $\Theta$, and $F_{ecl}$. We find that within the ranges tested, condensation temperatures between 1600K and 2200K can cause the observed $\Theta$ variations. Despite the brightness temperature of the planet being hotter than this range, we find that at the morning terminator the temperature can fall below these values, leading to increased cloud coverage. Outside this temperature range, clouds are either not present, or occur over most of the planet's dayside, resulting in a symmetric optical component and no negative $\Theta$ values in either case. An example grid is shown in Figure 4. In the case found requiring the smallest albedo change ($T_C$ =2000K, ε=8, κ=40), the entire range of $\Theta$ we observe, excluding the three outlier points in Figure 1 with $\Theta > 0.1$, can be explained by albedo shifts of only 0.05, although changes of order 0.1 combined with variation in $\varepsilon$ are more typical. This albedo shift produces variability in $F_{ecl}$ in the model of at most 25ppm, or

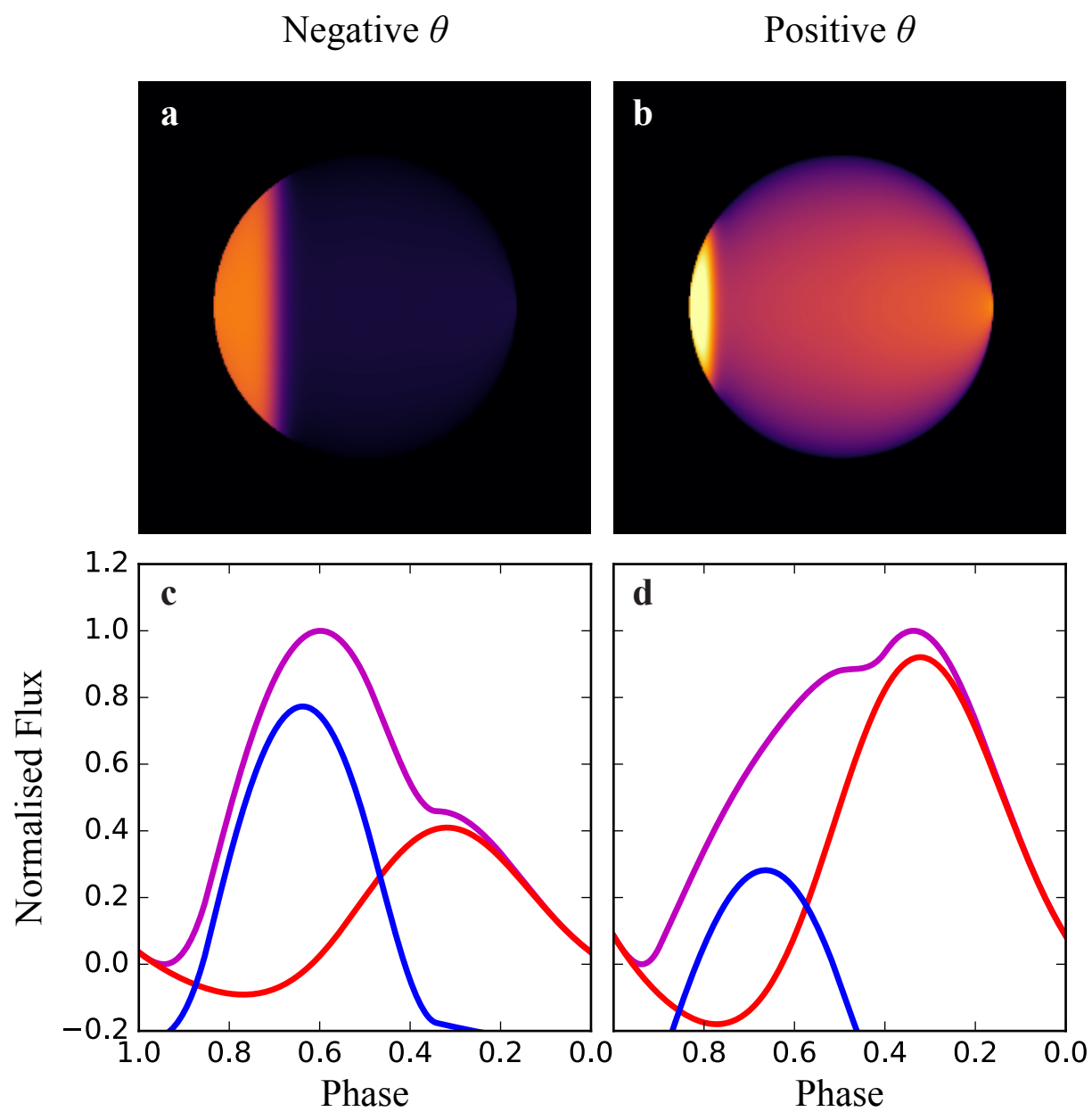

**Figure 3: Model outputs giving positive and negative $\Theta$ values.** a,b: Example flux maps of the planetary dayside. a,c: Increased cloud coverage results in an increased reflected flux and lower temperatures, leading to $\Theta$ = -0.1. b,d: Reduced cloud coverage results in lower reflected flux and increased temperatures, leading **to** $\Theta$ = 0.1. Variability in the wind speed bringing clouds from the morning terminator to the dayside could cause the varying cloud coverage. The temperature change at the substellar point between each case is of order 100K. Limb darkening is not shown here for clarity. c,d: The thermal emission (red), reflected flux (blue) and total (magenta) phase curve components associated with these $\Theta$ values, calculated using a published model. Phase curves are normalized to the total flux.

35% of the global fit value of 71.2ppm[6], smaller than the constraint of 51% we place above. At the above values of $T_C$ , ε, and κ, an albedo change of 0.1 corresponds to a peak temperature change of 100K on the planet. As the sharp edge of the Planck function is near the edge of the *Kepler* bandpass for HAT-P-7 b, this significantly changes the thermal flux emitted in the bandpass, and hence magnifies the change in observed peak offset. A wide range of parameter combinations can produce the observed values however. Several model grids for a range of temperatures are given in Supplementary Figure 6 and Supplementary Figure 7.

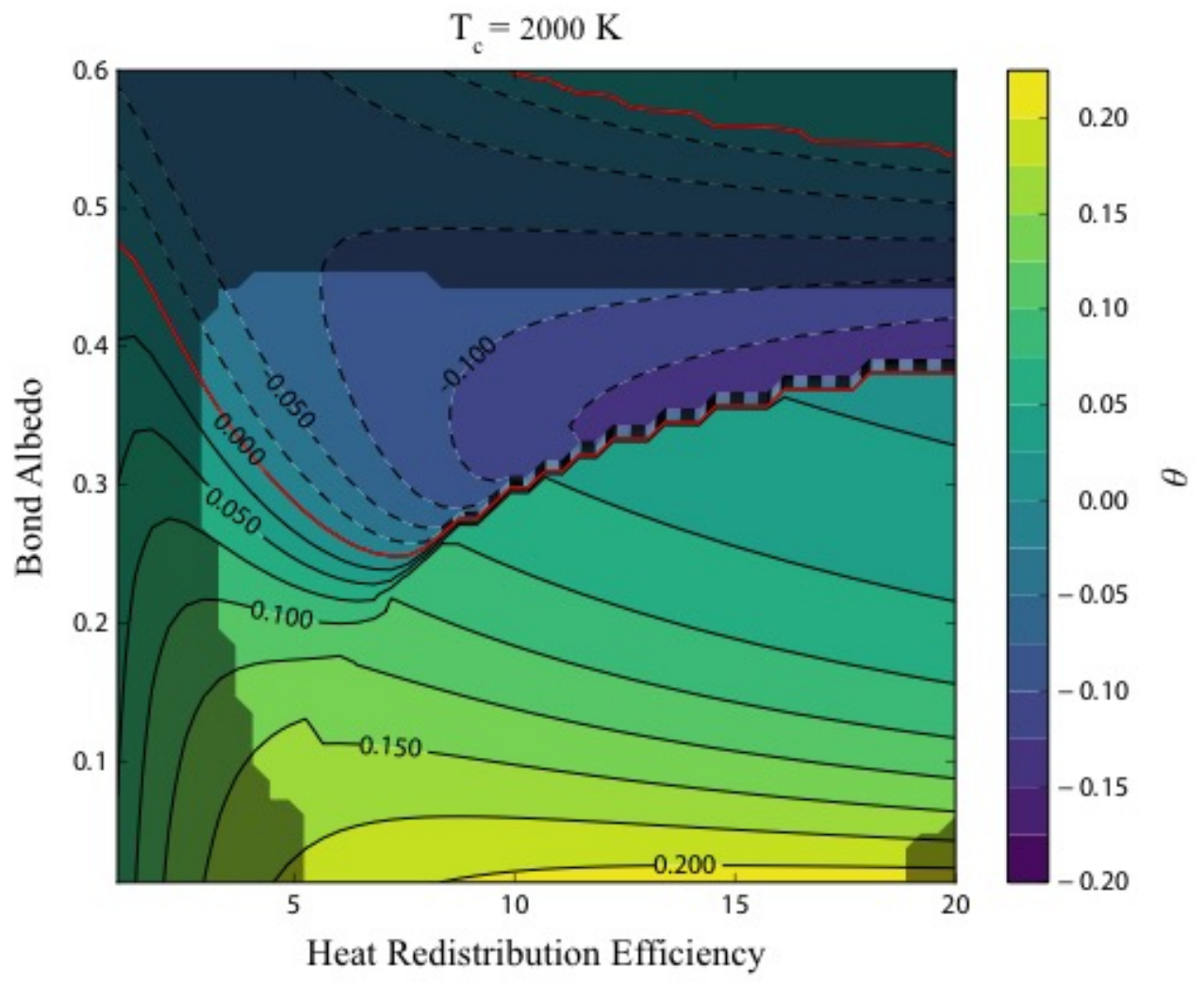

**Figure 4: Contour map of *Θ* variations.** $T_C$=2000K, $\kappa$=40, as a function of heat redistribution efficiency $\varepsilon$ and Bond Albedo $A_B$. Solid contours are positive; dashed are negative, red is zero. To produce the variations in *Θ*, the parameters must cross the red line repeatedly. The discontinuity seen at higher $\varepsilon$ values marks the boundary of cloud formation on the planet's dayside. The greyed out regions mark areas excluded by the +-59% limit on amplitude variations.

Here we discuss what physical mechanisms could lead to the above parameter changes, and hence the observed variations in *Θ*. Circulation models of tidally-locked hot Jupiters predict the presence of strong superrotating winds[15], which are responsible for the observed eastward shifts of thermal hotspots for hot Jupiters with infrared phase curves[14]. Whilst dayside temperatures of HAT-P-7 b as inferred from measured Spitzer transit depths (~2800 K) are likely to be too warm for possible condensates to form, the extremely fast winds can transport aerosols from the cooler nightside. Due to short horizontal advection timescales, aerosols may persist in regions between the morning terminator and the substellar point, despite the high dayside temperatures, before eventually evaporating towards the evening side of the planet.

Circulation models for planets like HAT-P-7 b with equilibrium temperatures of 2000-2200 K have been published[25], and suggest that condensate species $Al_2O_3$ (corundum) and $CaTiO_3$ (perovskite) would condense out at around 100 mbar on the nightside of these planets. According to the same set of models, these condensates could persist in the atmosphere at the morning terminator, and would then be advected onto the dayside. The cloud-covered proportion of the planet between the dayside and morning terminator will be dictated by the relative timescales of advection and evaporation. Increased wind speeds would reduce advection timescales and would allow the aerosols to travel further into the dayside atmosphere prior to evaporation; increased wind speeds would also move the thermal hotspot further towards the planet's evening terminator. A plausible mechanism for the phase offset variation in the Kepler lightcurves is therefore variation in the speed of the superrotating jet, which aperiodically transports a greater proportion of reflective aerosols onto the dayside and causes reflection from the cloud top to dominate the phase curve shape instead of the thermal contribution. The mechanism for such aperiodic variation in wind speed is unknown, but a possible driver may be the extreme tidal forces experienced by HAT-P-7 b due to its close orbit.

An increase in albedo on the dayside due to a larger concentration of advected aerosols will lead to a cooling of the upper atmosphere, which in turn will reduce the day-night temperature contrast and is therefore likely to weaken the superrotating flow[26]. This will act to reduce the transport of aerosols to the dayside and creates a feedback mechanism, resulting in oscillations in wind strength, dayside albedo and dayside temperature, explaining the observed phenomenon of the shifting phase curve peak.

Previous studies of the phase curve of HAT-P-7 b have ignored the time dimension, and show a wide range of often disagreeing values in measured amplitudes and derived albedos and temperatures, especially in the IR [5,6,9-13,27]. This has to date been put down to differing wavelengths, datasets and analysis methods. We show here that the planet itself is variable, and care should be taken when simultaneously analysing measurements of Hot Jupiters at disparate times. This detection of variations in an exoplanet's atmosphere implies that the temporal dimension of planetary weather can now be explored outside the solar system. Given the wide variety of exoplanet types observed, including many not seen in the solar system, new models will be required to explain dynamical atmospheric changes. Future space missions such as CHEOPS, JWST, PLATO and ARIEL will be able to expand on this data and create a sample of temporally resolved exoplanet atmospheres.

# Requests and Correspondence

Correspondence and requests for materials should be addressed to DJA at d.j.armstrong@warwick.ac.uk.

# Acknowledgements

DJA acknowledges funding from the European Union Seventh Framework programme (FP7/2007- 2013) under grant agreement No. 313014 (ETAEARTH). EdM acknowledges support from the Michael West Fellowship. JBa is funded by the ERC 'ExoLights' project. This paper includes data collected by the Kepler mission. Funding for the Kepler mission is provided by the NASA Science Mission directorate. All of the data presented in this paper were obtained from the Mikulski Archive for Space Telescopes (MAST). STScI is operated by the Association of Universities for Research in Astronomy, Inc., under NASA contract NAS5-26555. Support for MAST for non-HST data is provided by the NASA Office of Space Science via grant NNX09AF08G and by other grants and contracts.

# Author Contributions

DJA obtained and detrended the data, developed and fit the phase curve models, implemented the atmospheric model, produced the figures and wrote the manuscript. EdM developed the discussion, contributed to the tests performed to check the results, and tested the results with his own models. HPO contributed to the phase curve model, and produced visual interpretations of the results. JBa developed the discussion of the atmospheric processes behind the peak offset variations. JBl provided the initial development of the phase curve model. NFS contributed to development of the figures. All authors commented on the manuscript.

# Author Information

# METHODS

**Data Source**

We obtained publicly available data from the NASA *Kepler* satellite, which operated from 02-May-2009 to 11-May-2013[8]. The spacecraft reorients itself every ~90 days, separating data into 'Quarters'. We downloaded the complete dataset for HAT-P-7 (Kepler-2, KIC 10666592), comprising Quarters 0-17 in both long (29.5 minute, data release 24 and 25) and short (58.9 second, data release 24) cadences. There is a known problem with the short cadence data smear correction in data release 24, and so with short cadence data release 25 data not yet available we focus our analysis on the long cadence data. Each data release and cadence gives consistent results. Data Quarters were stitched together by normalizing each quarter by its own median flux, producing a single lightcurve.

**Data Detrending**

We use several differing methods of detrending, to reduce the chance that our chosen method affects our results. We start with the raw SAP_FLUX data. Long term trends in this data are removed through fitting a 3$^{rd}$ degree polynomial function. First points within 0.05 in phase of transit centre are masked, using published ephemeris[6]. The data is then split into sections of width 0.3d. A running window around each of these sections is used to fit a 3$^{rd}$ degree polynomial. We also trial a 2$^{nd}$ degree polynomial; although the resulting lightcurve contained increased systematic noise there was no significant effect on the fit results. The polynomial is fit iteratively, with datapoints discrepant from the previous fit by more than 8 sigma masked from the succeeding fit. This is repeated for 10 iterations. We trial several different lengths of window for fitting, comprising 10d, 5 planetary orbital periods (11.02d), and 3 planetary orbital periods (6.61d). Each produces consistent results. Shorter windows give decreased systematic error in the lightcurve, at increased risk of overfitting. As we found no significant difference in the resulting fit parameters between window sizes, we use 3 planetary periods for the final analysis. The window is not allowed to cross gaps in the data of greater than 1d. In cases where such gaps are found, the window is extended before or after the gap to meet the window size required. The resulting polynomial is then divided out from the section under consideration, and the process repeated for each section. At this stage, data within 0.7d of any gap larger than 1d is removed, as data ramps and other systematics are prevalent near gaps in the *Kepler* data. We further manually remove any regions showing clear instrumental systematics. The *Kepler* data before and after detrending are shown in Supplementary Figure 8.
As an alternative detrending method, we employ the Covariance Basis Vectors (CBVs) provided at the Michulski Archive for Space Telescopes (MAST). These are used to remove instrumental trends common to nearby stars on the CCD. We use the first 5 CBVs for detrending, enacted through the kepcotrend tool in the PyKE software package[28]. Planetary transits and occultations are masked from the CBV fit. Following the CBV step, the polynomial fit is repeated as above in order to flatten the lightcurve, as using CBVs alone leaves significant long term trends present. The results from the CBV detrended data are consistent. We also tried fitting to the lightcurve produced by the CBVs alone, with no polynomial step. The resulting lightcurve contained significant instrumental noise visible by eye, however we found fitting results consistent with the polynomial lightcurves. This strengthens our confidence in the analysis and detrending applied.
Finally we compared the results between the long and short cadence data. These were again consistent, aside from a 27 day region centred on 543 (BJD-2454833), where a small but systematic offset was found. We conservatively removed data from this region before analysis.

**Phase curve fitting**

The full optical curve of HAT-P-7 b is shown in Supplementary Figure 9. We model the planetary phase curve using a previously published model[6]. This has 4 key components – the planetary transits/occultations, the tidal bulge caused by the planet's mass on the star, Doppler beaming from the star's reflex motion, and variation from the planetary brightness itself. Each is significant for this system. Also included is a cosine 3$^{rd}$ harmonic term empirically found to be present[6]. A full list of fit parameters and detailed description of the model is given in the Supplementary Information. We hold all parameters associated with ellipsoidal variation, Doppler beaming, and the planetary transit constant at the values of [6], which are listed in Supplementary Table 1. The planetary brightness amplitude, peak offset and secondary eclipse depth are allowed to vary, along with a constant offset in flux, leading to a 4 free-parameter fit.
To test the variation of the phase curve with time, the lightcurve is split into predefined segments of planetary orbits. We test segment sizes ranging from 5 orbits up to 20 orbits. Segments are not allowed to cross quarter boundaries. Segments with less than half the expected number of datapoints (calculated from the segment size and the data cadence) are not included. Each segment is then phase folded using the planetary ephemeris and fit independently. We remove data during planetary transit to avoid this affecting the fit. We also trial a sliding window with the same size, shifting the window by one orbit and refitting. This allows higher time resolution, but does not produce independent fits, as some of the same data is included in successive fits. As such the discrete segment fits should be used for measuring variation significance, but the sliding window can be used to explore the variation in greater detail.
We derive errors on these fits through a combination of 3 separate methods. These are the covariance matrix, the 'prayer-bead' residual permutation technique (e.g. ref [23]), which is designed to incorporate systematic noise into the errors, and error resampling. These are explained in more detail in the below section. The method giving the largest error is used for each fit and parameter.
For final values we use a segment size of 10 planetary orbits. This allows robust detection of the planetary brightness in each segment, at 9 sigma confidence on average. Decreasing the segment size to 5 planetary orbits shows no clear shorter timescale variation while reducing the significance of each fit. Values for all fits are given in a supplementary file, and the individual fits and segments for discrete fits shown in Supplementary Figure 2.
It would be ideal to fit for the planetary mass dependent processes simultaneously with the phase curve parameters. However, we note that the planetary mass must stay constant in time. As such, fitting for it would require a combined fit across the whole dataset, while variations were allowed independently in each measured segment (i.e. an at least 250 parameter simultaneous fit). This is computationally expensive to perform, and so we test the impact of the planetary mass by using the RV derived mass, as it is independent of the mass derived from the phase curve. We test fitting to the phase curve segments using different fixed planetary masses, at the $-1\sigma$, best fitting and $+1\sigma$ values of the RV derived mass (1.725, 1.781, and $1.862M_J$ respectively[6]). The resulting change in phase curve fit parameters was at the level of 5% of the parameter errors, and hence we ignore this effect. We conclude that the effect of an imperfectly known planetary mass, through Doppler beaming or ellipsoidal variation, is insignificant in this case.

We cut fits for which the phase curve amplitude was not detected at $3\sigma$ confidence, or where the detected amplitude was lower than 30ppm. In these cases, the peak offset is not well defined. This removed 26 fits from the original 592 produced using the sliding bin.

**Fitting and Error methods**
We combine two methods of error derivation (prayer-bead and error resampling) when calculating our values and errors. These errors are then compared to the error derived through the covariance matrix, and the largest error used in the final values.
The prayer-bead technique (otherwise known as residual permutation) proceeds as follows. The data is initially fit, and initial best fit parameters obtained. The residuals of the data from this initial fit are stored. These residuals are then permuted; shifted through the data in time such that correlated noise in the residuals is maintained, but occurs at different phases. At each step, the shifted residuals are applied to the initial best fit model, then the resulting data refit. This can be repeated for as many datapoints as are available – beyond this number the fits repeat each other. We combine this method with another well-known method, error resampling. Here multiple fits are repeated, but at each fit each datapoint is resampled from a Gaussian distribution formed from its own errors. We also randomise the initial guesses for each fit parameters, drawn from a uniform distribution within 20% of their initial best fitting values. To combine both methods simultaneously, we shift the residuals by a random amount at each iteration. With these variations applied each time, the fit is repeated for 1000 iterations, for each segment of lightcurve that we consider. In each iteration, the best fit parameters are found through least squares minimisation, enacted using the scipy optimize.leastsq routine in Python[29]. The final values and errors are found from the distribution of resulting fit parameters. We extract the $15.87^{th}$, $50^{th}$, and $84.14^{th}$ percentiles from this distribution, to form the lower 1 sigma, best fit value, and upper 1 sigma errors respectively. In this way the errors contain 68.27% of the distribution. We then test the resulting errors against those derived from taking the square root of the diagonal of the covariance matrix. While the covariance matrix errors are general smaller than those derived through the resampling or prayer-bead methods, if they are not we adopt the larger error for each data segment.

**MCMC Fitting**

We also test the fitting using an MCMC routine. This is applied to two representative phasecurve segments, one each with significant positive and negative $\Theta$. We implement an MCMC around the phase curve model using the open-source emcee software[30]. We fit for the original 4 parameters, plus an additional factor f which is multiplied to the datapoint errors. This is to account for correlated noise not present in the given errors. We applied the MCMC to two representative phase curve chunks, at times 226 and 1238 (BJD-2454833). These have significantly high and low $\theta$ respectively. The MCMC was performed with 100 walkers, started with randomised parameters near the same initial guess as the previous fitting methods. We iterated the chain for 5000 steps, finding that all parameters had converged by 1000 steps. The remaining 4000 values (i.e. 400000 iterations considering all walkers) were used to test for parameter correlations and consistency with the other fitting. Triangle plots and derived values are shown in Supplementary Figure 4 and Figure 5. Values for all parameters matched those of the previous analysis, although we find that the MCMC errors are underestimated compared to the prayer bead analysis. The larger errors are used for final values.

**Quarter to Quarter variations**

It has previously been shown that the transit depths of HAT-P-7 b vary on a seasonal cycle, as the target falls on different CCD modules each Quarter[31]. Such systematic variations could bias our results. However, the variation in transit depth is at the level of 1% around the average, much less than the variation we detect. As further evidence that our observed variability is not caused by seasonal variations, we note that the observed variability does not repeat on a yearly timescale (i.e. when the star falls on the same CCD module), and shows no significant discontinuities between Quarters, as seen in the transit depth analysis[31]. We also note that no significant power is detected in the periodogram at multiples of the 90d Quarter timespan (see below).

**Orbit to Orbit variations**

In planetary brightness amplitude, peak offset and secondary eclipse depth we sometimes see apparently significant variations on planetary orbital timescales. These should be treated with caution, as the phase curve is not significantly detected in individual orbits of the planet. We cannot exclude that they are due to either low level stellar activity, or small scale changes in the data arising from the specific method used, although they may be the signal of short timescale variations in the planetary atmosphere.

**Periodicity Analysis**

A Lomb-Scargle[32,33] periodogram analysis of $\Theta$ (using discrete bins and hence independent fit parameters) returns no significant frequencies. We estimated significance by running periodograms on 10000 simulated white noise datasets, with the same time sampling and scatter as the observed data. For each simulated dataset, the periodogram peak was extracted. The 1 and 3$\sigma$ percentiles of the distribution of maximum peaks were used as significance thresholds. We find that peaks in the real data are all weaker than 1$\sigma$ under this significance test, and hence concluded that no periodicity is detected.

**Exclusion of systematics/stellar variability**

At the high level of precision at which we are interested, there is potential for systematic and instrumental noise in the *Kepler* data to affect our results. Although HAT-P-7 is an inactive star[34], low-level stellar variability may also have an effect, and gravity darkening has been noted in the transit[35]. Firstly, all phases in transit are cut during the analysis, thereby removing any impact from gravity darkening. The 3$^{rd}$ order harmonic previously found[6] has an amplitude of 1.93ppm on average, and is hence negligible to the overall phasecurve, and undetectable in the lightcurve segments we analyse. We trial a large array of detrending methods in an attempt to test for issues caused by our choice of methods. We find the results robust to any combination of parameters which we trial. Futhermore, non-planetary variability would have to occur in a way which mimicked a planetary phase curve, and would also need to occur on a much shorter timescale than the typical Quarter-to-Quarter systematics which *Kepler* presents. Variability with timescale longer than 3 planetary periods is removed by the polynomial detrending, and variability with timescale shorter than 10 planetary periods is partially averaged out by the binning of 10 successive phase curves. The stellar rotation period is >13.23 days[36], and hence would be removed by the polynomial detrending. Even before polynomial detrending, no significant starspot activity (which occurs on the stellar rotation period) has been detected by previous studies[34]. Shorter timescale stellar variability (such as `flicker'[37]) is averaged out through combining multiple planetary orbital periods in each fit. Finally, we note that our `sliding window' fits present a smooth variation in fit parameters, which would not be the case if single events (such as cosmic rays, detector malfunctions, stellar flares, or one-off instrumental effects) were causing the variation. Any such one off event would produce sharp variations in the fit parameters as the event entered the window and as it left.

We further test for systematics by selecting random planetary orbits from throughout the dataset, then performing the fit on the combination of these. If the signals we see are coherent, as they should be if they are connected to the underlying atmospheric dynamics, then randomizing the orbits used should remove the variability. We find that when using 10 planetary orbits the variability is markedly reduced, with the standard deviation of the extracted $\Theta$ values dropping by ~20%. We tested that this reduction in variability was reasonable for a real signal by considering a simulated dataset with similar properties to the true variation in $\Theta$. The simulated data consisted of two sinusoids with periods of 10 and 100 days (a simple recreation of a coherent variability signal and some correlated noise), and amplitude 0.035, overlaid on a white noise signal with $\sigma$=0.1. Considering randomized binned samples from the simulated dataset as opposed to consecutive bins led to a drop in the standard deviation of $\Theta$ of ~20%, matching the effect on real data. Using 20 planetary orbits rather than 10 removed the variation entirely.

**Injection Testing**

As a further test against systematic noise, we identify two nearby stars with similar magnitude and temperature to HAT-P-7, observed on the same *Kepler* CCD module. These are KIC10861893 and 11027406. For each of these targets, we follow the same procedure to get the lightcurve as for HAT-P-7. We inject a constant phase curve using the HAT-P-7 b parameters from [6] into the lightcurves before the polynomial flattening stage. We then run the same detrending and fitting code and attempt to extract phase curve parameters. In both targets, no significant variation is found in $\Theta$ (Supplementary Figure 3), although some structure can be seen, likely as a result of underlying starspot activity. For KIC10861893 however, marginally significant variation is seen in $A_p$, on the same level as for HAT-P-7 b. As the $A_p$ variation in HAT-P-7 b was inconclusive, it appears that this systematic noise source was included in our errors, as we would expect. This test confirms that the variations in $A_p$ do not originate with the planet. The lack of variation in $\Theta$ supports the physical origin of these variations in HAT-P-7 b.

**Alternative Explanations**

It is worth exploring whether other physical effects could give rise to the observed variation. A stellar rotation-planet orbit resonance, combined with variable stellar activity, is excluded in the previous section. To give rise to a signal occurring on the planetary orbital period, other potential origins for the signal are necessarily related to the planet. They could include a variable dark spot on the star, consistently below the planet's position. This would require some form of magnetic connection between the planet and star, and hence a particularly strong planetary magnetic field. Such a spot would be expected to affect the transit depths of the planet, yet does not, making this explanation unlikely. Similarly, a variable bright spot on the opposite side of the star to the planet could lead to the observed signal, although it is unclear why such a spot would exist. Both explanations are intrinsically unlikely, as due to the planet's misalignment with the stellar spin maintaining a coherent interaction between the two would be difficult.

**Atmospheric Model**

We implement a published semi-analytical model to allow us to explore the atmosphere of the planet[24]. This model takes certain atmospheric parameters and calculates the expected phase curve. These parameters are:

$A_B$: Bond Albedo. Controls the proportion of light reflected by the planet. This reflected light comes from both cloudy and non-cloudy regions of the atmosphere, and hence the albedo in part controls the cloud coverage.

f: Greenhouse factor. A free scaling factor to represent possible greenhouse effects. As we do not have enough information to constrain this, we set it to unity, corresponding to no greenhouse effect.

$\varepsilon$: Heat Redistribution Factor. Defined as the ratio between the radiative timescale and the advective timescale[38]. $|\varepsilon|>>1$ corresponds to efficient heat redistribution, such that temperature is smoothed out around the planet's longitude by strong winds. $|\varepsilon|<<1$ represents the opposite, where heat is radiated faster than it is transported, and longitudinal differences in temperature will be large. We assume positive values for $\varepsilon$, which correspond to superrotating winds on the planet.

$\kappa$: Cloud Reflectivity Boosting Factor. The factor by which cloudy regions are brighter than non-cloudy regions in reflected light.

$T_C$: Condensation Temperature. $A_B$ and $\varepsilon$ together with the known planetary and stellar parameters allow for the calculation of a temperature map on the planet. Where the temperature falls below $T_C$, clouds form. For positive $\varepsilon$, cloud formation occurs preferentially on the morning side of the planet, as the hottest point is shifted to the afternoon side. With a temperature map, it is possible to calculate the thermal contribution to the phase curve, by integrating the Planck function over the *Kepler* bandpass then integrating the local emission over the planetary disk. The reflected light contribution can be calculated by considering the cloudy and cloud free regions. We refer the interested reader to the original publication[24] for full details.

## Condensates

In the main text we propose corundum or perovskite as plausible condensates in HAT-P-7 b's atmosphere. As such it is important to check if these condensates may have sufficient material to form the necessary clouds. Based on the work of Wakeford et al. (2016) following the relative elemental mass calculations of Lodders (2010), we find that if corundum and perovskite are the sole condensables for Al and Ti respectively then over one scale height the optical depth in the Kepler bandpass would be ~2-4 for corundum and ~0.1-1 for perovskite. If the cloud extends for greater than one scale height these depths could increase. As such there is sufficient available material to form an optically thick cloud for corundum but less so for perovskite, mainly because Al is more abundant than Ti. As there are a number of assumptions here this does not exclude perovskite, but makes corundum the more likely of the two. We consider only particles of 0.1 and 1 micron in size, which are strongly scattering at short wavelengths for both species, and any cloud formed would be likely to be reflective.

## Visualisation

We visualise the changes on the planet through extracting temperature maps and cloud coverage from the atmospheric model. The model is not fully constrained, and so we arbitrarily select a region of parameter space which produces the observed variations with modest parameter changes (0.38-0.17 in $A_B$, 12.9-5.3 in $\varepsilon$, with $\kappa$=40 and $T_C$=2000K). We extract the thermal and reflected flux for each contour shown in Figure 4 between these values. For each observed $\Theta$, we interpolate between these contours to produce a representative flux map, cloud coverage and brightness. These are shown in the linked video, which gives a visual representation of the changes observed. We stress that this video is for clarification and explanatory purpose only; the values used to produce it are not best fits. Nevertheless, the proportion of thermal emission and reflected light seen at each epoch is the necessary combination to produce the directly observed peak offset.

In the video the bright band at the left (morning) side of the planet shows the area of cloud formation, and associated reflected light. The more diffuse bright spot towards the right (afternoon) side of the planet represents the thermal emission seen in the *Kepler* bandpass. Note that while the thermal emission observed changes significantly, this only corresponds to relatively small changes in the underlying temperature. This is due to two effects. Firstly, the shorter wavelength edge of the thermal emission overlaps with the edge of the *Kepler* bandpass. As such, a slight increase in temperature markedly increases the amount of flux emitted into the *Kepler* bandpass. This would be expected to increase the secondary eclipse depth and phase curve amplitude, and indeed this is found by the model. However, the increase in depth and amplitude is not significant enough to be seen over our (relatively large) errors on these parameters, as described in the main text. Secondly, we do not show the effects of planetary limb darkening in the video, which would decrease the flux from the planetary limbs. Displaying the planet without limb darkening amplifies the cloud and thermal variations to increase clarity.

## Data Availability Statement

Individual fit values are provided in the supplementary information as a csv file. Any other data that support the plots within this paper and other findings of this study are available from the corresponding author upon reasonable request.

## Further References

# Variability in the Atmosphere of the Hot Giant Planet HAT-P-7 b

## Supplementary Information

### Tables

| Parameter | Unit | Value |
|---|---|---|
| **System** | | |
| KIC | - | 10666592 |
| P | d | 2.2047354±0.0000001 |
| $T_0$ | BJD†-2454833 | 121.358572±0.000004 |
| $a/R_*$ | - | $4.1545^{+0.0029}_{-0.0025}$ |
| i | degrees | $83.143^{+0.023}_{-0.020}$ |
| **Star** | | |
| $T_*$ | K | 6350±80 |
| $R_*$ | $R_{sun}$ | $1.84^{+0.23}_{-0.11}$ |
| $M_*$ | $M_{sun}$ | $1.47^{+0.08}_{-0.05}$ |
| $u_1$ | - | $0.3497^{+0.0026}_{-0.0035}$ |
| $u_2$ | - | $0.1741^{+0.0057}_{-0.0044}$ |
| **Planet** | | |
| $R_p$ | $R_J$ | $1.419^{+0.178}_{-0.085}$ |
| $R_p/R_*$ | - | $0.077524^{+0.000017}_{-0.000022}$ |
| $M_p$ | $M_J$ | 1.63±0.13 |
| $A_3$ | ppm | -1.93±0.23 |
| $\Theta_3$ | phase | $0.0163^{+0.0054}_{-0.0052}$ |
| **Doppler/Ellipsoidal** | | |
| $\alpha_1$ | - | 0.0657 |
| $\alpha_2$ | - | 1.22 |
| $\alpha_d$ | - | 3.41 |

Table 1: HAT-P-7 Input Parameters[6]. † BJD=Barycentric Julian Date.

# Figures

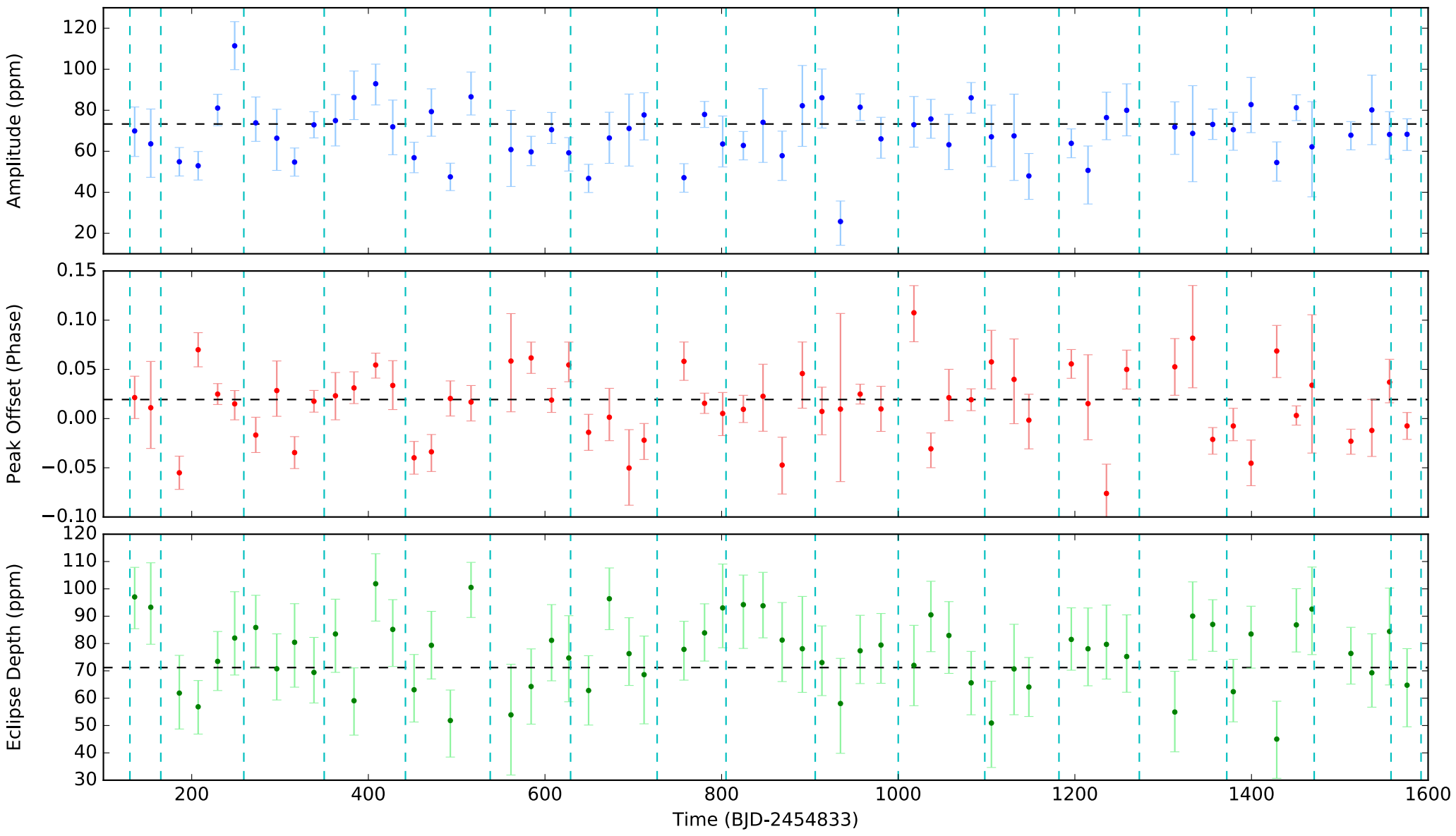

Supplementary Figure 1: Independent best fitting parameters. Variation in peak amplitude (top, blue), peak offset (middle, red) and secondary eclipse depth (bottom, green). Best fit parameters resulting from fits to discrete lightcurve segments are shown, resulting in fewer, but independent, points than the convolution method leading to Figure 1. The Kepler quarter boundaries are shown as vertical dashed lines. Error bars are 1σ errors on the fit parameters.

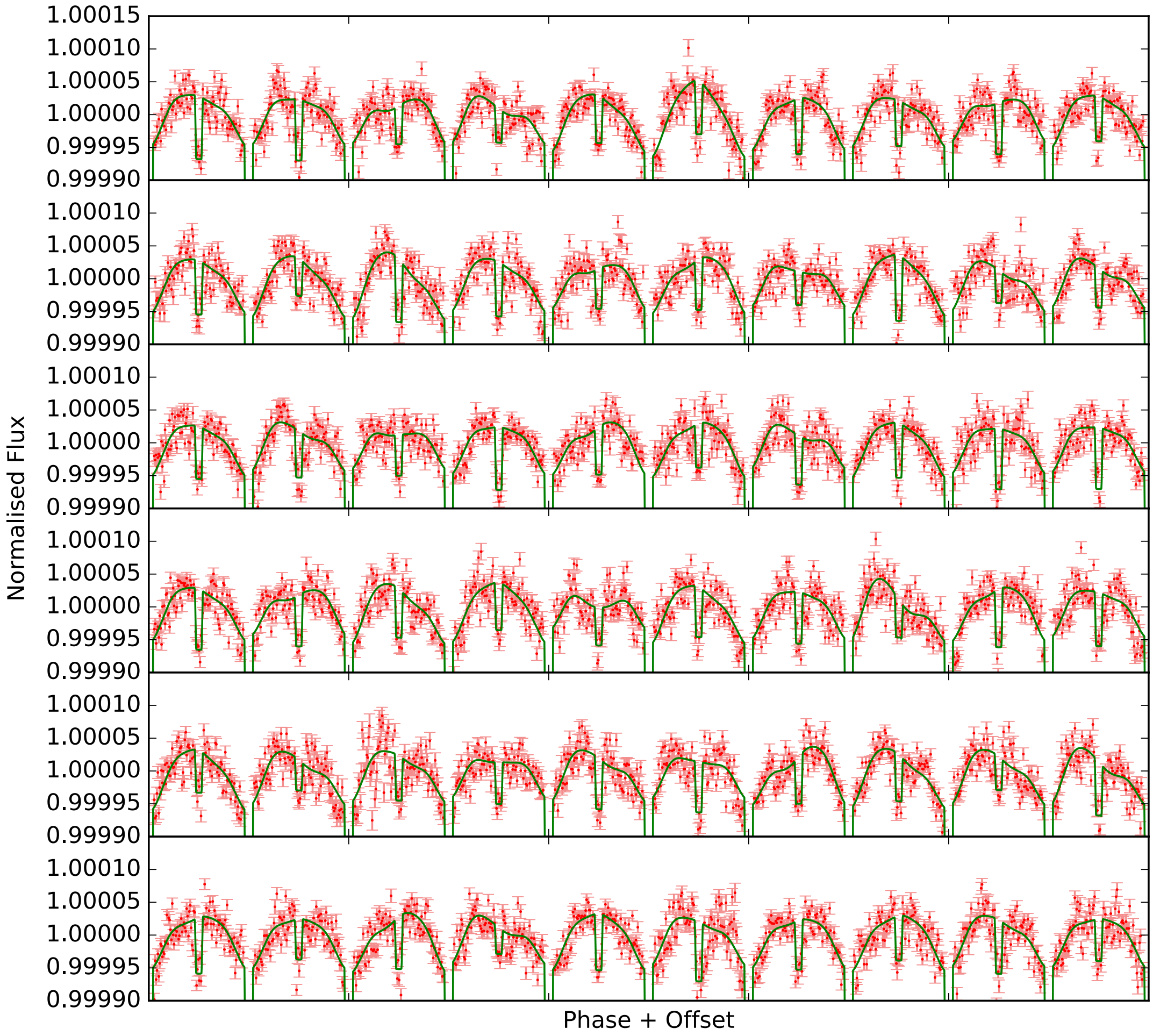

Supplementary Figure 2: The data segments used to derive the fit parameters of Supplementary Figure 1. The best fit models are plotted in green. Reading left to right and top to bottom, they correspond to each datapoint shown in that Figure. Each segment has been binned to 100 bins for clarity. Error bars are the standard error on the mean of the datapoints contributing to each bin.

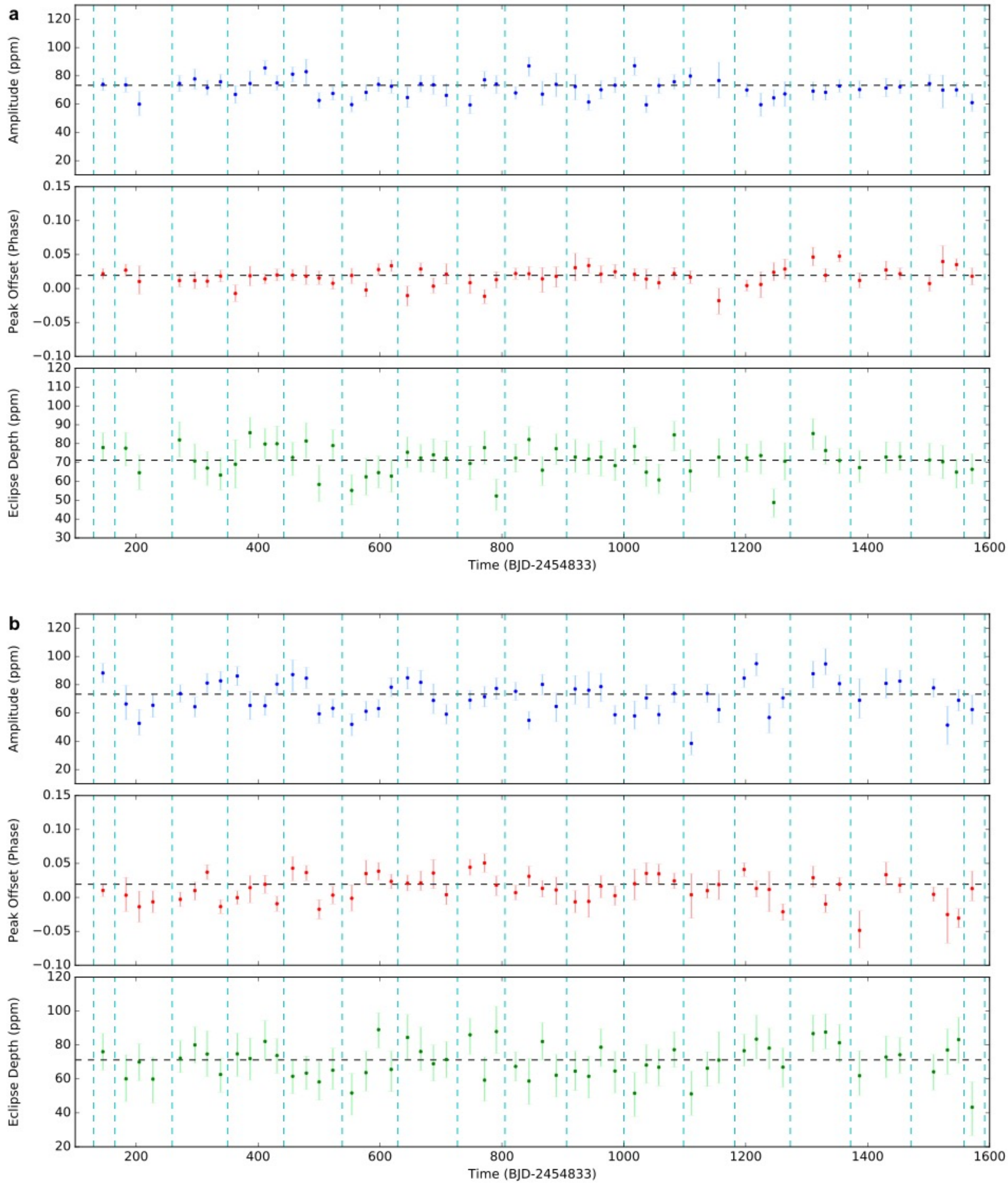

Supplementary Figure 3: Best fitting parameters from injection tests. Panel a: As Supplementary Figure 1 for KIC11027406 after injection with a constant HAT-P-7 b phase curve. Panel b: As Supplementary Figure 1, for KIC10861893 after injection with a constant HAT-P-7 b phase curve. Marginally significant variation in $A_p$ is seen for KIC10861893, implying the presence of systematic noise biasing the fits. The variations in $\Theta$, while showing some structure, are on a much smaller scale than those observed in HAT-P-7 b. Error bars are $1\sigma$ errors on the fit parameters.

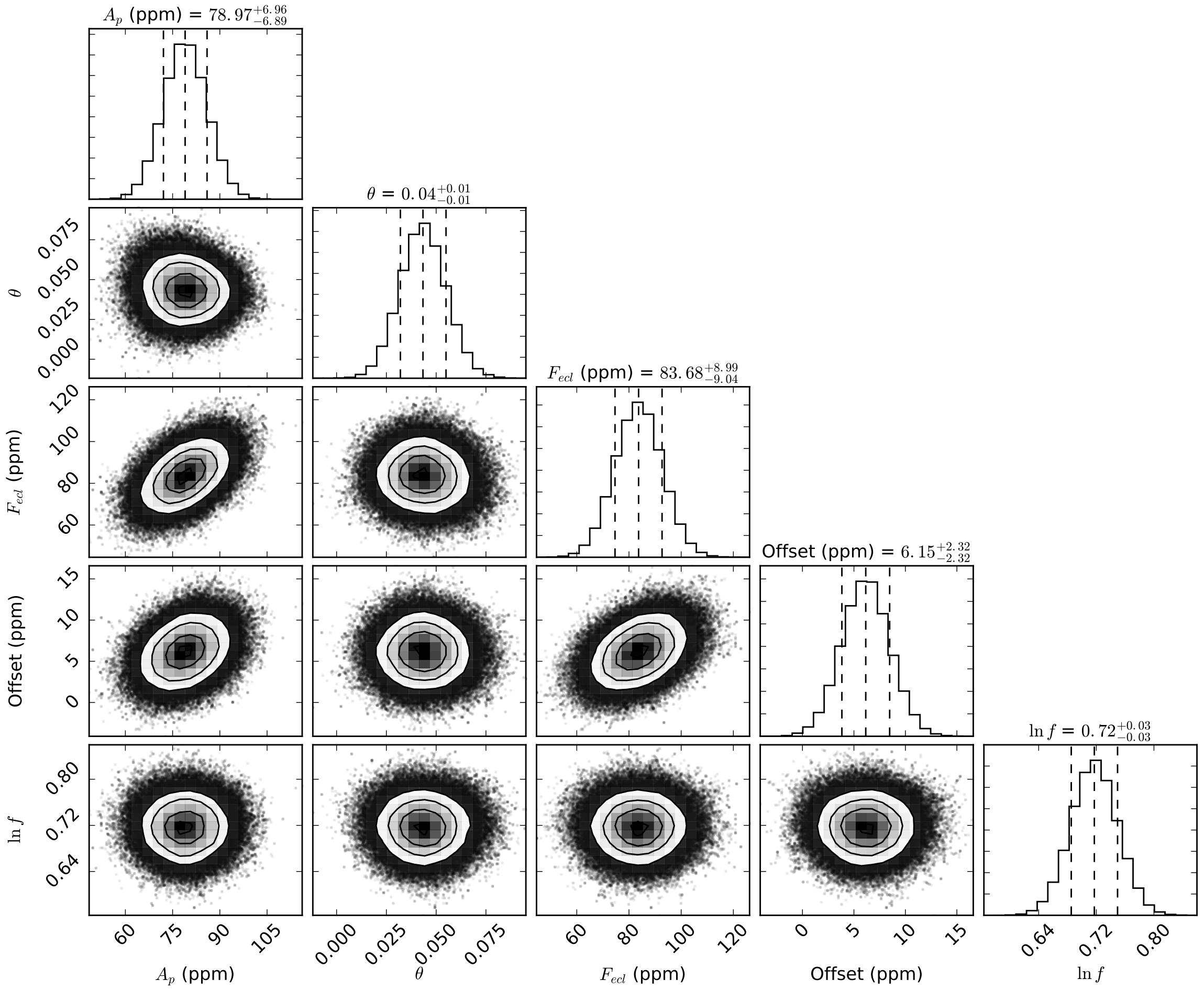

Supplementary Figure 4: MCMC parameter correlations. Triangle plot of an MCMC run applied to a phase curve segment at time 226 (BJD-2454833), with significant positive Θ. The outputs are consistent with those of the prayer bead analysis, although the errors from the MCMC are slightly smaller. The larger errors are used for final values. An additional parameter, f, is fit for which scales up the errors by a factor of ~2, to account for correlated noise not present in the given errors.

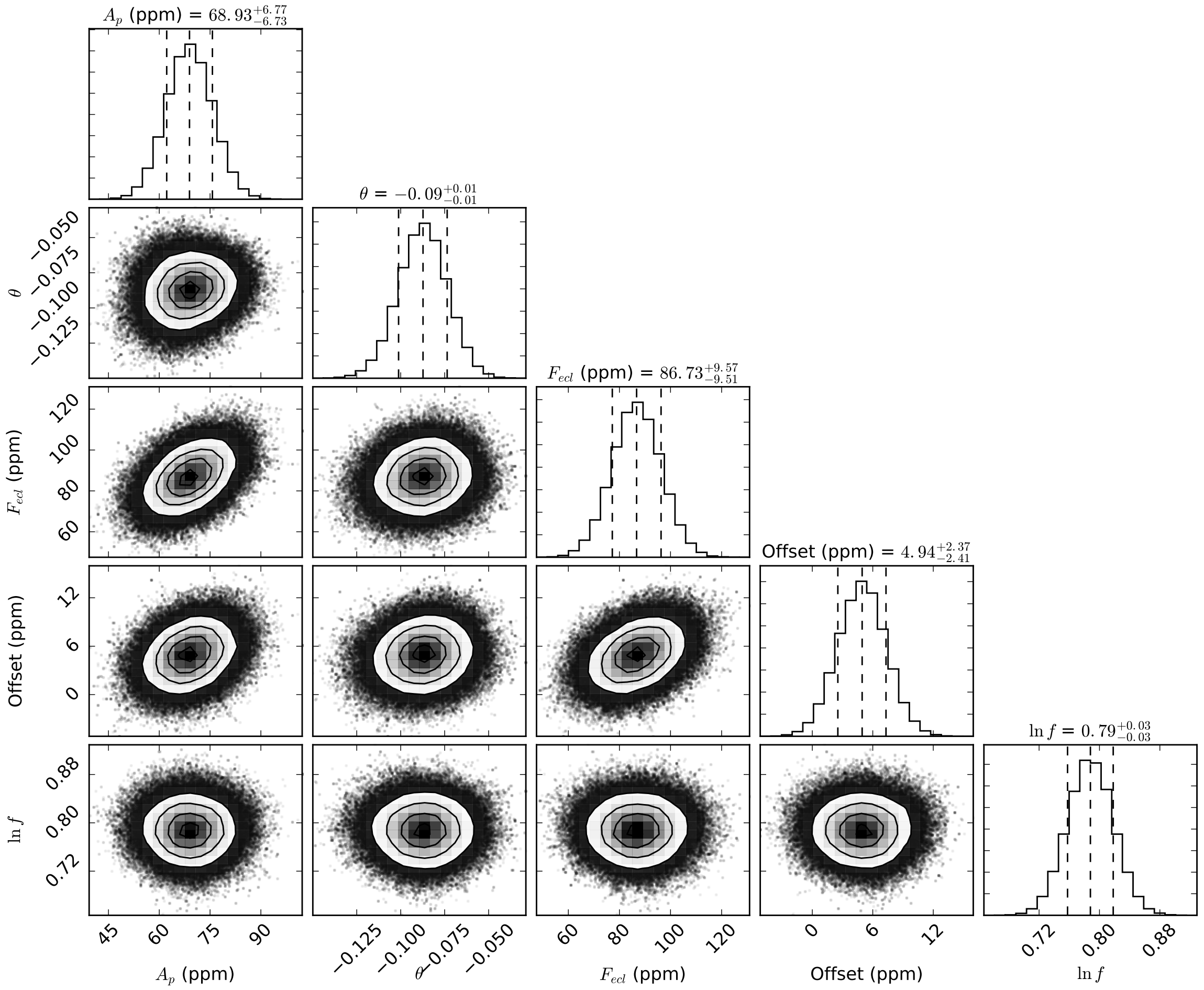

Supplementary Figure 5: MCMC parameter correlations. As Supplementary Figure 4 for a phase curve segment at time 1238 (BJD-2454833). This segment has significant negative Θ. The parameters are again consistent with the prayer bead results, with the MCMC analysis giving slightly smaller errors.

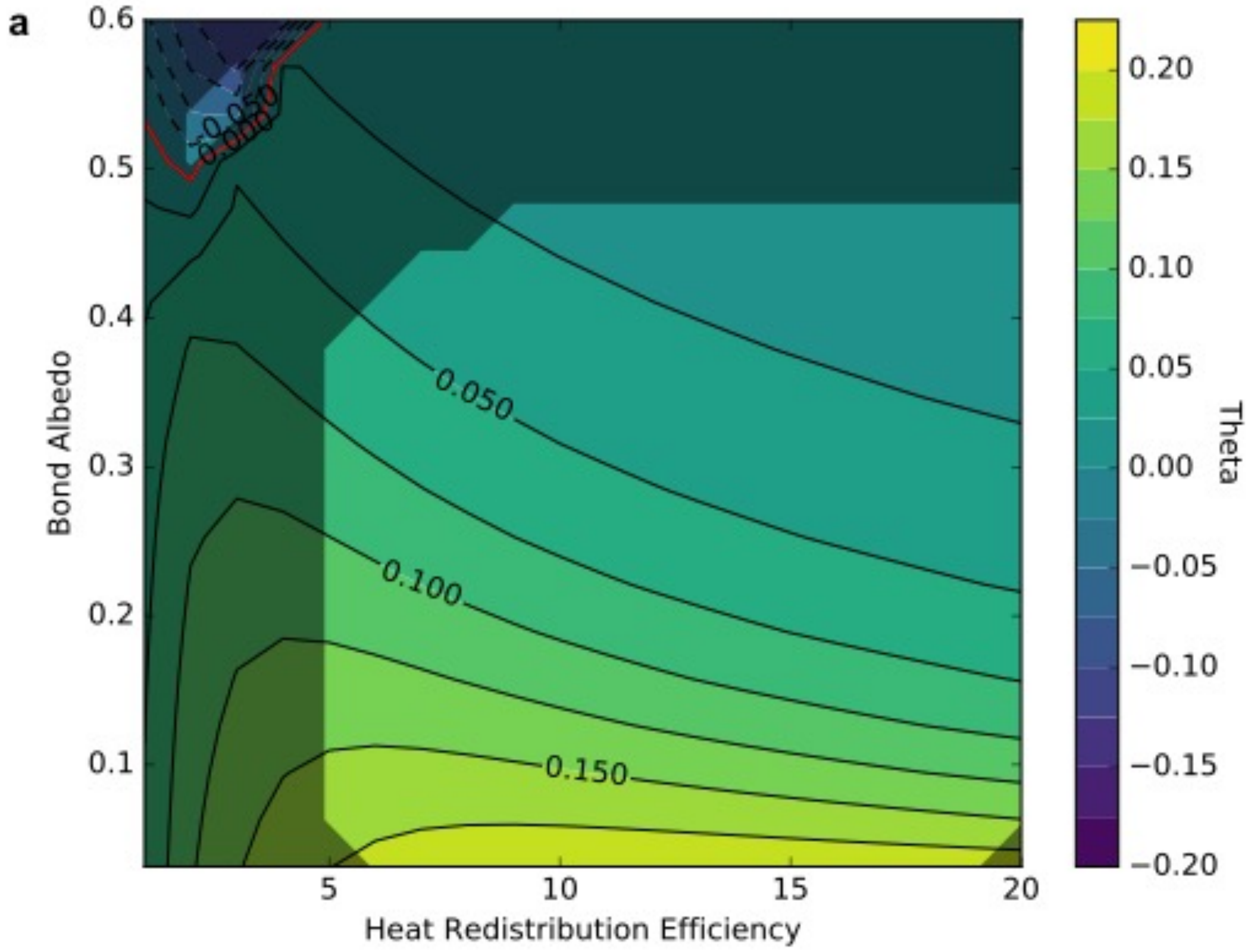

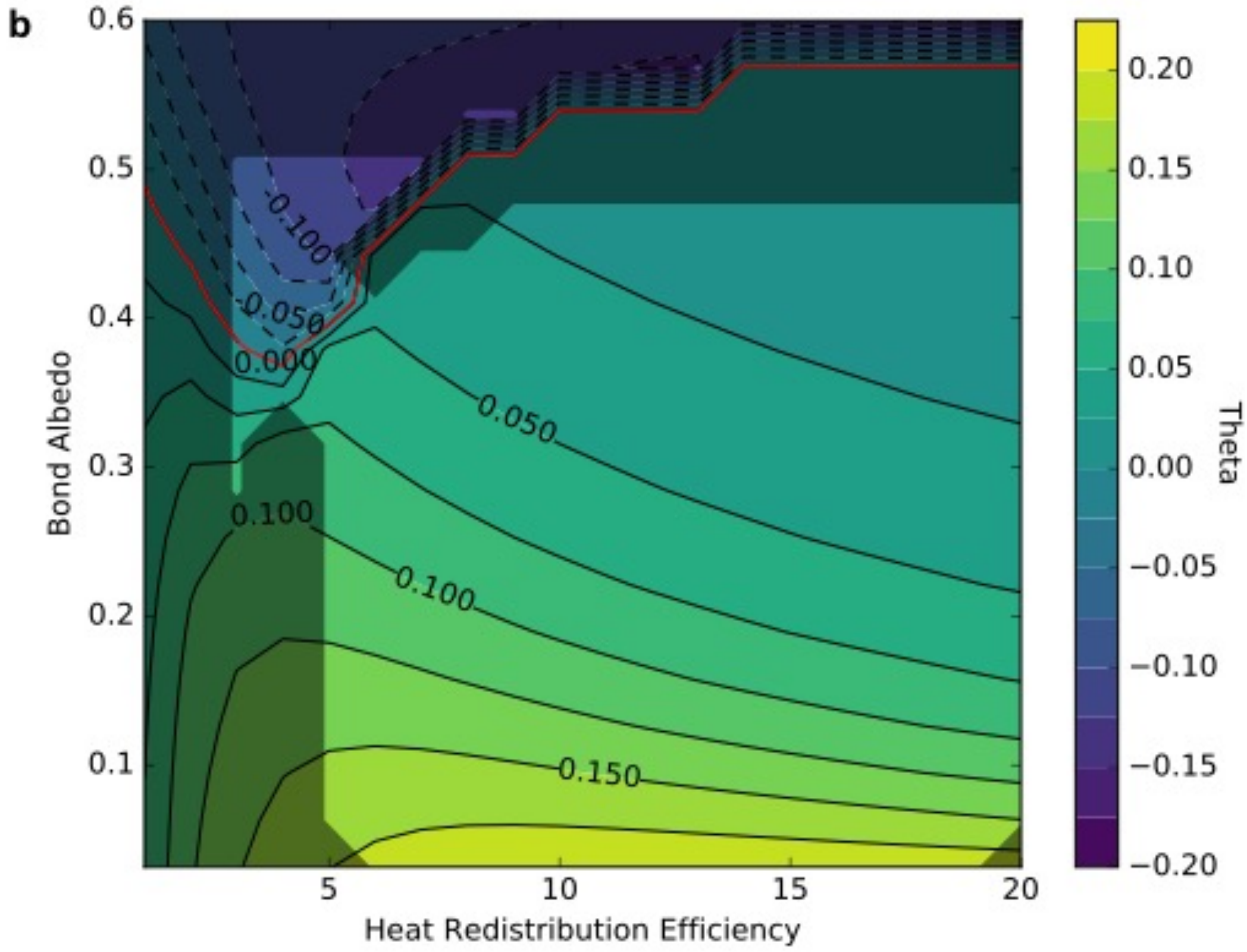

Supplementary Figure 6: Peak offset predicted by model. As Figure 4 for $T_C = 1600K$ (top), $T_C = 1800K$ (bottom) and κ=40. Reducing the value of κ makes negative contours less deep, but does not change the structure of the grid. Producing all the observed values of Θ requires transitioning through a regime disallowed by the observed amplitude, but is possible.

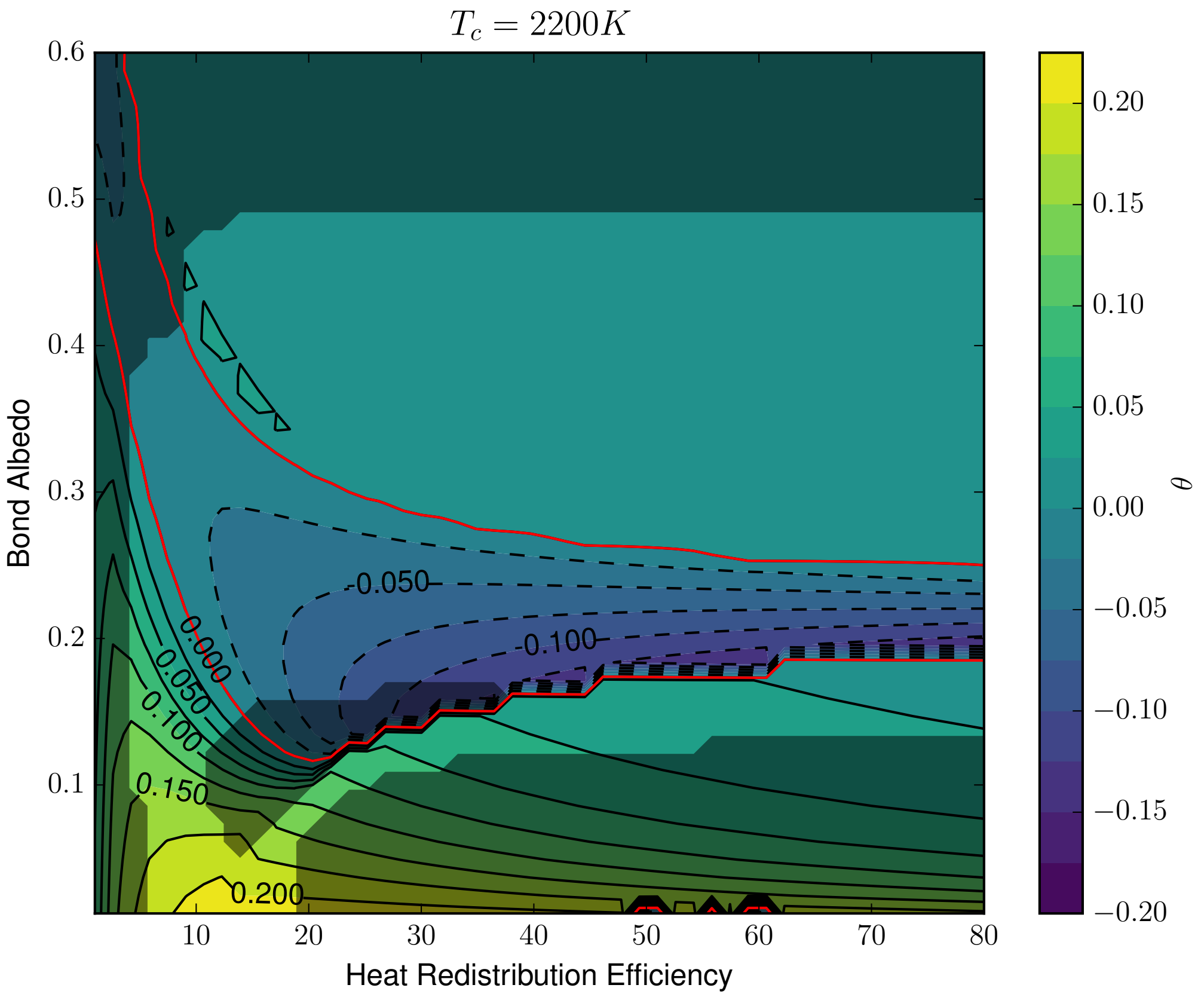

Supplementary Figure 7: Peak offset predicted by model. As Figure 4 for $T_C$ = 2200K and κ=40. Reducing the value of κ makes negative contours less deep, but does not change the structure of the grid. The grid has been extended up to ε=80, as it is not possible to achieve the maximum negative offset in the previously used ε=[0,20] grid range.

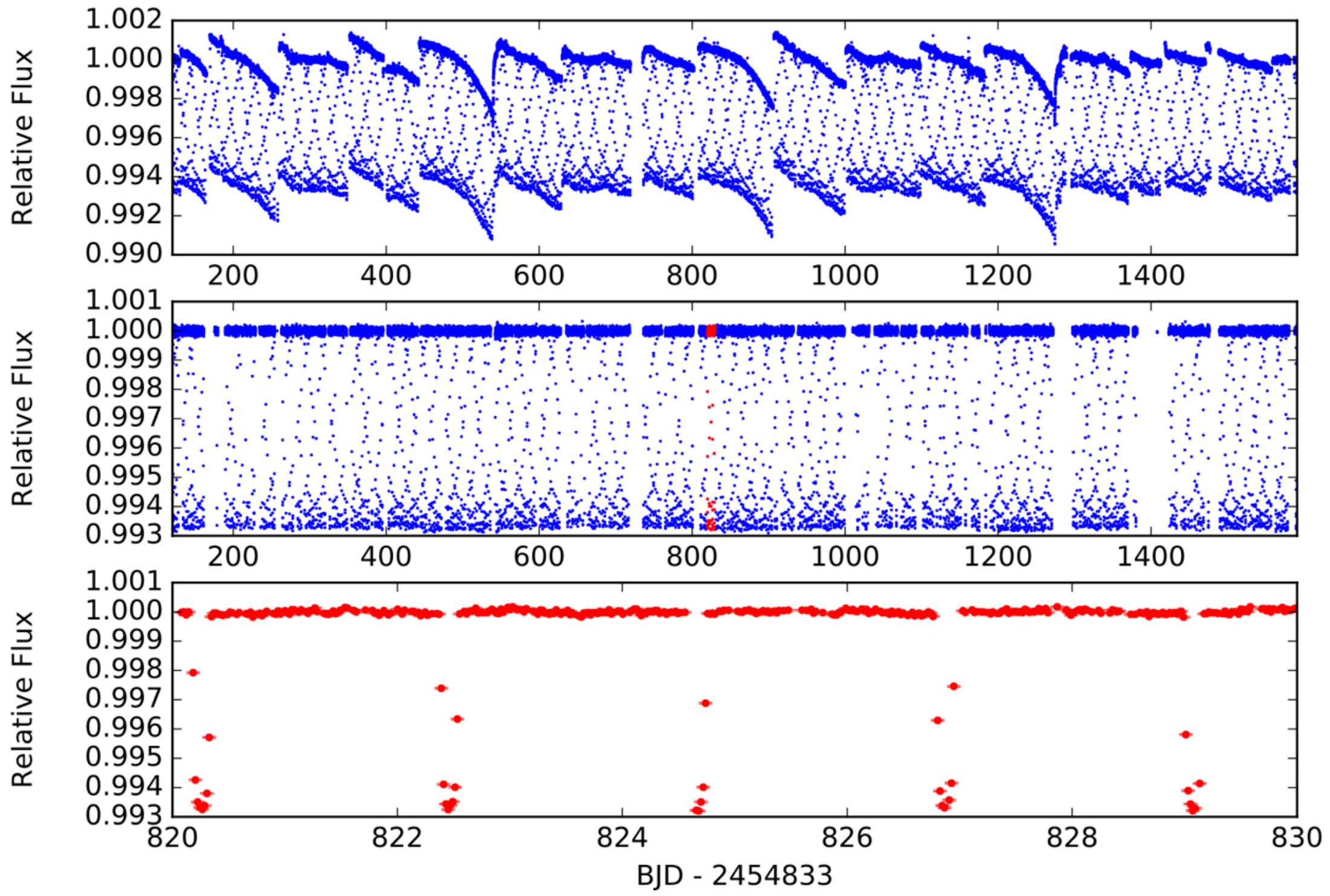

Supplementary Figure 8: Lightcurve detrending process. Kepler data of HAT-P-7 before detrending (top), after detrending (middle) and zoomed in to show 4 planetary orbits (bottom).

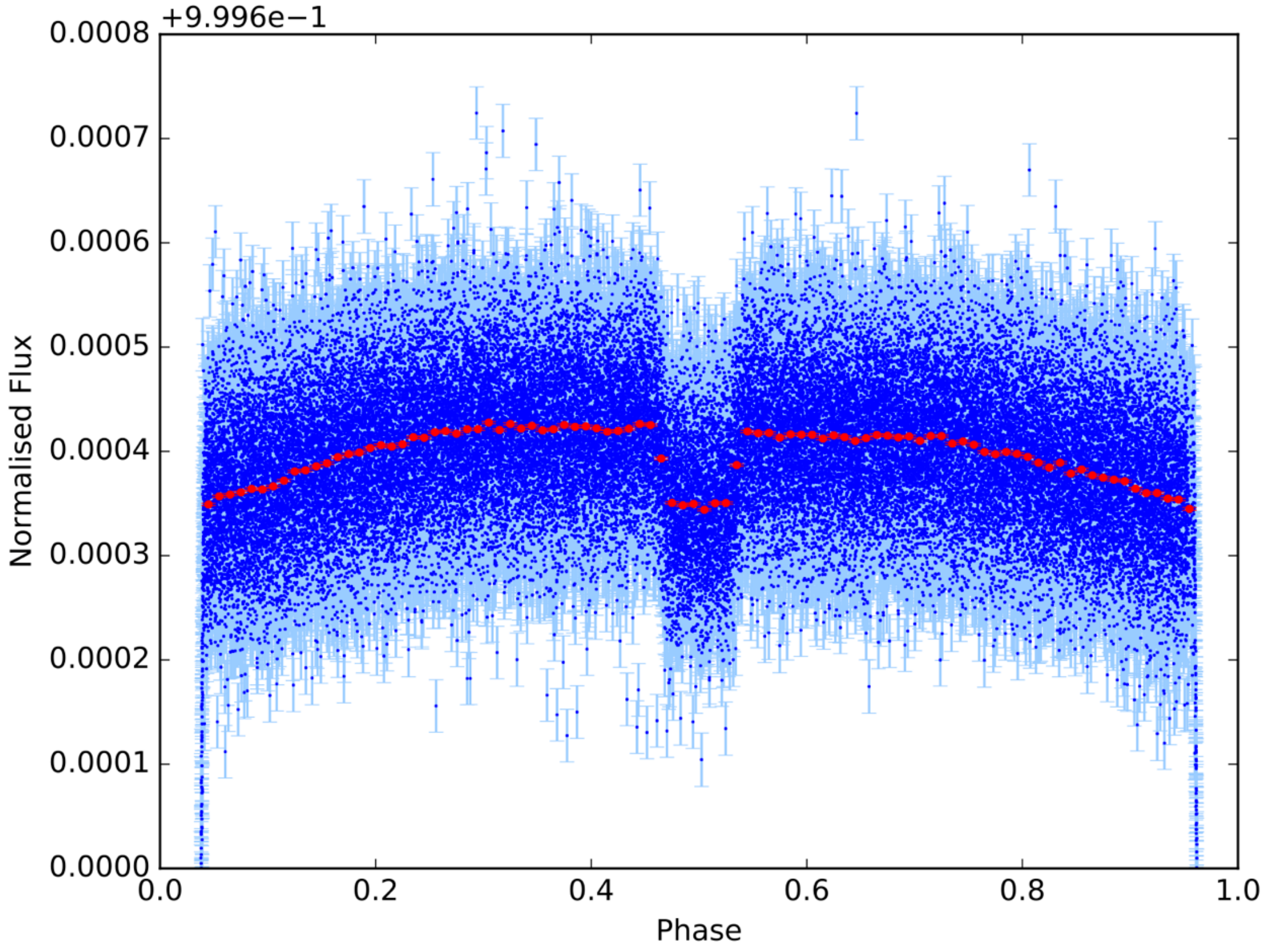

Supplementary Figure 9: HAT-P-7 b phase curve, showing the entire dataset. The unbinned data are in blue, and 100 bins shown in red. The planetary transit is at phase 0 and is below the scale of the graph. Error bars are the 1σ photometric errors on each datapoint.

**Phase curve model**

We use the model of [6]. This is summarised here for clarity. As HAT-P-7 b has no detectable eccentricity [9,10], we assume circular orbits throughout.

The model is comprised of five components: a quadratically limb darkened Mandel-Agol transit model $F_{\mathrm{transit}}$ [41], using the parameters given in Supplementary Table 1 (although note that datapoints in transit were removed before fitting), contributions from the planetary mass $F_m$, the secondary eclipse $F_{ecl}$, the planetary brightness $F_p$, a cosine third harmonic of the planet's period $F_3$, and an offset $C$. The combined flux at phase $\phi$ is given by

$$F(\phi) = F_{\mathrm{transit}}(\phi)F_m + F_{\mathrm{ecl}}(\phi) + F_p(\phi + \theta) + F_3(\phi - \theta_3) + C$$

with

$$\phi = 2\pi \frac{t - T_0}{P}$$

where $T_0$ represents the first mid-eclipse time, and $P$ the planetary orbital period.

The planetary brightness $F_p$ is defined by a Lambert sphere, with

$$F_p = A_p \frac{\sin z + (\pi - z)\cos z}{\pi}$$
$$\cos z = -\sin i \cos(2\pi(\phi + \theta))$$

where $A_p$ is the peak amplitude, $\theta$ is the phase offset in peak brightness, and $i$ is the planetary orbital inclination.

The secondary eclipse is modelled as a second Mandel-Agol transit, for the occultation of a uniform source with the planetary parameters given in Supplementary Table 1, with depth $F_{ecl}$. $F_{ecl}$ and the above $F_p$ component are allowed to vary as a function of time, and are refit for each subsection of the lightcurve considered.

The planetary mass contributes through both Doppler boosting and ellipsoidal variation. Doppler boosting arises from the changing radial velocity of the host star as the planet orbits. When the star is moving towards the observer, the stellar flux is increased via Doppler boosting and also blueshifted, resulting in a changing contribution within the *Kepler* bandpass. The reverse occurs when the star moves away from the observer. Ellipsoidal variation results due to changes in the visible surface area of the star. The stellar surface has a tidal bulge associated with the planet – as the planet orbits, this bulge moves into and out of visibility, changing the visible flux. The planetary mass contribution is given by

$$F_m = M_p \left\{ \left(\frac{2\pi G}{P}\right)^{1/3} \frac{\alpha_d \sin i}{c M_*^{2/3}} f_d - \frac{\alpha_2 \sin^2 i}{M_*} f_e \right\}$$

where $M_p$ is the planet's mass, $G$ the gravitational constant, $M_*$ the stellar mass, $c$ the speed of light, and $\alpha_d$ the photon-weighted bandpass-integrated beaming factor. The values of $\alpha_d$ and $\alpha_2$ are given in Supplementary Table 1, and are taken from [6]. $f_d$ and $f_e$ are the phase dependent components of the Doppler boosting and Ellipsoidal signals respectively. They are given by

$$f_d = \sin(2\pi\phi)$$

$$f_e = \left(\frac{a}{R_*}\right)^{-3} \cos(4\pi\phi) + \left(\frac{a}{R_*}\right)^{-4} f_1 \cos(2\pi\phi) + \left(\frac{a}{R_*}\right)^{-4} f_2 \cos(6\pi\phi)$$

with $f_1$ and $f_2$ determine the amplitude of higher order ellipsoidal variations, and are given by

$$f_1 = 3\alpha_1 \frac{5\sin^2 i - 4}{\sin i}$$

$$f_2 = 5\alpha_1 \sin i$$

The value for $\alpha_1$ is given in Supplementary Table 1, and is as calculated in [6].

Lastly, the cosine third harmonic is described by

$$F_3 = A_3 \cos\big(6\pi(\phi - \theta_3)\big)$$

with $A_3$ the third harmonic amplitude and $\theta_3$ the phase offset of the signal. This third harmonic signal was discovered in [6], and has an unknown origin. We include it with their parameters, and hold it constant with time.

Of all the parameters mentioned here, most are held constant at their values fit from the whole dataset. The parameters allowed to vary between lightcurve segments are $A_p$, $\theta$, $F_{ecl}$, and $C$, the latter to allow for small shifts in flux between segments. We hold the remaining parameters constant to simplify the model, and because their effect is expected to be negligible on the phase curve itself. We discuss those parameters which could impact the phase curve here:

Orbital Period: The published value and error used[6] lead to an error of less than one *Kepler* cadence in phasing over the entire dataset, and less over each phase curve chunk. Hence we consider the effect negligible.

Epoch: As above with the orbital period, the published value and error lead to an effect of less than one *Kepler* cadence.

Planet Mass: See Methods for a test of the effect of fixing the mass, which led to no significant effect.

Third Harmonic: This harmonic was detected at a level of 2ppm[6], at 8.4σ significance using the entire dataset. We are unable to detect the harmonic in a 10 orbital period phase curve chunk. As such, while the harmonic is included in the model, fitting for it simultaneously would have negligible effect.

Radii Ratio: The primary transit is removed before fitting. However, the planet to star radius ratio affects the shape of ingress and egress of the secondary eclipse. Given the constraints found on secondary eclipse depth variation (51% at 3σ), the shape of ingress or egress of the secondary eclipse is not well determined enough to have any significant effect.

## Supplementary Information References